\documentclass[usegraphicx,usenatbib]{mn2e}
\title[Major Mergers of Galaxy Haloes]{Major Mergers of Galaxy Haloes:
Cuspy or Cored Inner Density Profile?}
 
\author[M. Boylan--Kolchin and 
C.--P. Ma]{Michael~Boylan--Kolchin$^{1,2}$\thanks{E-mail:
mrbk@astro.berkeley.edu} and Chung--Pei~Ma$^2$\thanks{E-mail:
cpma@astro.berkeley.edu} \\
$^1$Department of Physics, University of California, Berkeley, CA
94720, USA\\
$^2$Department of Astronomy, University of California, Berkeley, CA,
94720, USA}
\begin{document}

\pagerange{\pageref{firstpage}--\pageref{lastpage}} \pubyear{2003}

\maketitle

\label{firstpage}

\begin{abstract}
We present the results from a series of collisionless $N$-body
simulations of major mergers of galaxy dark matter haloes with density
profiles having either inner cusps or cores.  Our simulations range
from $2 \times 10^5$ to $10^7$ particles, allowing us to probe the
phase-space distribution of dark matter particles in the innermost
regions (less than 0.005 virial radii) of cold dark matter haloes, a
subject of much recent debate.  We find that a major merger of two
cored haloes yields a cored halo and does not result in a cuspy
profile seen in many cosmological simulations.  This result is
unchanged if we consider mergers with parent mass ratios of 3:1
instead of 1:1. Mergers of a cuspy halo with either a cored 
halo or a second cuspy halo of equal mass, on the other hand, produce
cuspy haloes with a slightly reduced inner logarithmic slope.  Cuspy
haloes, once formed, therefore appear resilient to major mergers.  We
find the velocity structure of the remnants to be mildly anisotropic,
with a Maxwellian velocity distribution near the centre but not in the
outer portions of the final haloes.  Violent relaxation is effective
only during the early phase of mergers, with phase mixing likely to be
the dominant relaxation process at late times.
\end{abstract}


\section{Introduction} 

In the currently favored cosmological constant plus cold dark matter
($\Lambda$CDM) model of cosmology, structure forms hierarchically:
small dark matter haloes collapse at high redshifts, while larger and
more massive objects form through a series of minor mergers that
accrete smaller mass haloes and major mergers with comparable mass
haloes.  Since dark matter haloes provide the gravitational potential
wells necessary for galaxy formation \citep{wr78}, a detailed
understanding of the effects of mergers is necessary in order to test
models of galaxy formation and evolution within the $\Lambda$CDM
paradigm.

Recent numerical studies indicate that the hierarchical build-up
process results in haloes with significant substructure populations:
hundreds to thousands of subhaloes to the smallest resolvable mass
scales appear to reside within galaxy sized haloes, comprising
approximately 10\% of a halo's mass \citep{ghigna98, klypin99,
moore99b}.  A number of recent papers have investigated the evolution
and detailed properties of the substructure using both semi-analytic
models and numerical simulations.  Depending on factors such as the
relative mass, pericentric distance, and halo concentration, the
orbits and masses of the subhaloes can be strongly affected by
dynamical friction and tidal stripping (e.g. \citealt{vw99, tb01,
hayashi03, taffoni03, benson03}).  Dense subhaloes that survive the
tidal process can sink to the centre of the parent halo, impacting its
inner structure (e.g. \citealt{dekel03,ddh03}).

In contrast to minor mergers, where large haloes swallow up smaller
ones in a relatively gentle fashion, major mergers are more violent
events involving systems of approximately equal mass that can be
studied in detail with numerical simulations only.  The interest in
major mergers has a long history.  \citet{toomre} noted that many
observed features of galaxy pairs can be explained by tidal
interactions that could promote merging.  They further speculated that
if such interactions were common enough, elliptical galaxies might
naturally be explained as the merger remnants of spiral galaxies.
This suggestion has led to numerous studies of galaxy merging and has
been quite successful in helping explain the origin of ellipticals.
In some of the earliest simulations, \citet{white78,white79}
studied several types of mergers using 250 simulation particles per
halo and concluded that products of mergers of cored galaxies were
significantly more centrally concentrated than their progenitors, with
more extended haloes having an approximate power-law density profiles
$\rho \propto r^{-3}$ that resembled the de Vaucouleurs' empirical
profile for the surface-brightness of ellipticals \citep{dv48}.  The
higher resolution simulations by \citet{pearce93} with $\sim 8000$
particles per halo also exhibited the $r^{-3}$ behavior with evidence for
homology.
Simulations with multiple components representing dark matter haloes,
stellar disks, and stellar bulges have been a source of much
theoretical progress (e.g. \citealt{bh92} and references therein) and
are essential for elucidating the dynamical coupling between the dark
matter and stellar components in galaxies and for comparing simulation
results with observations. Pure dark halo simulations, however, are
required for a controlled study of the non-baryonic dark matter that
dominates the gravitational potential of seemingly all galactic
systems.

In this paper we perform high-resolution $N$-body simulations of
mergers of two equal-mass dark matter haloes and study how the
phase-space distribution of dark matter particles is affected by such
violent events and how the merger product relaxes into equilibrium.
We pay particular attention to the central regions of the haloes
and examine whether the initial inner structure of a dark matter halo,
either with a flat core or a sharp cusp, is disrupted or preserved by
major collisions.  Our study complements the numerous recent
cosmological $N$-body studies of the properties and evolution of dark
matter haloes, most of which have reached the somewhat surprising
conclusion that over a wide range of mass scales, these haloes 
are universally cuspy and have radial density profiles with
$\rho(r) \propto r^{-\gamma}$ (with $1.0 < \gamma < 1.5$) in the
central region and $\rho \propto r^{-3}$ near the virial radius
\citep{nfw97, fm97, moore99a, ghigna00, fkm03}.  The seemingly generic
prediction of an inner cusp agrees with earlier simulations
(e.g. \citealt{dc91}) but is contrary to the inferred dark matter
distribution in some dwarf and low surface brightness galaxies
(e.g. \citealt{mdb98, simon03}, though also see \citealt{swaters03})
as well as in spiral galaxies \citep{salucci} and has lead to
speculation that the $\Lambda$CDM model is in crisis. 

The simulations of major mergers performed in this paper provide a set
of controlled experiments for us to address questions such as: How
does the phase space density of dark matter haloes evolve and
re-equilibrate during major mergers?  Is the cuspy profile an
attractor region of the phase space so that merger remnants of two
non-cuspy (i.e. cored) haloes will end up cuspy?  Do cuspy haloes,
once formed (by any process), retain their
inner structure after inevitable major mergers with other haloes?
Unlike cosmological simulations that are designed to incorporate many
complicated processes at once, the simulations performed here allow us
to separate out the effects of minor mergers and to focus on how major
collisions
change the spatial and velocity structures of any initial haloes of
our choice. Additionally, cosmological parameters are unimportant for
the simulations themselves since major mergers are local processes of
two fully collapsed objects.  In this case, all we need is the initial
phase space distribution of each galaxy halo and the initial relative
positions and velocities of the haloes themselves.  By simulating
collisions of two galaxy haloes with high spatial resolution and up to
5 million particles (for each halo), we are able to follow their
phase-space evolution and the density profiles of the remnants down to
$\sim 0.004$ $r_{200}$ in our highest resolution simulation.

Section 2 of this paper describes the initial setup and
the parameters used in the numerical simulations.  We also discuss
in some detail the tests performed to quantify the numerical effects
of two-body relaxation on the central regions of the haloes and the
technical issue of how to determine the centre of a merged halo.
Section 3 contains our results from the ten production runs for
the energy, density profiles, velocity distributions, and phase space
structure of merger remnants.  Section 4 discusses the energy
distribution and the relaxation of a merged halo to an equilibrium
state, with an emphasis on violent relaxation and phase
mixing. Section 5 contains a summary of our results.

\section{Procedure}    
\subsection{Initial Conditions}

We study two general types of inner density profiles for galaxy haloes:
cored and cuspy.  For the cored profile, we use an isothermal sphere
with a core radius and an exponential cut-off \citep{hern93}:
\begin{equation}
    \rho(r)= \frac{1}{2\pi^{3/2}} \frac{\alpha M}{r_{\rm cut}}
	\frac{\exp(-r^2/r_{\rm cut}^2)} {r^2+ r_{\rm core}^2}
\label{hernquist}
\end{equation}
where $M$ is the halo's mass, $\alpha$ is a normalization constant
given in terms of $x \equiv r_{\mathrm{cut}}/r_{\mathrm{core}}$ as
$\alpha^{-1} =1-\sqrt{\pi}x e^{x^2}\mathrm{erfc}(x)$,
$r_{\rm cut}$ is the outer exponential fall-off radius and $r_{\rm
core}$ is the core radius.  For the cuspy haloes, we use the Navarro,
Frenk, and White (NFW) profile \citep{nfw97}:
\begin{equation}
      \rho(r)= \frac{\rho_c \bar{\delta}}{(r/r_s) (1+ r/r_s)^2}
\end{equation}
with scale radius $r_s$ and characteristic density $\bar{\delta}$ given by
\begin{eqnarray}
   r_s= \frac{r_{200}}{c} = \frac{1}{c} \left(\frac{3M_{200}}{800 \pi \rho_c}
       \right)^{1/3}\\
   \bar{\delta}=\frac{200c^3} {3[\ln(1+c)-c/(1+c)]}
\end{eqnarray}
where $c$ is the halo concentration parameter and $r_{200}$ is the
radius at which the average interior density is 200 times the critical
density of the universe $\rho_c=3H^2/8 \pi G$.  (We also adopt
$r_{200}$ as a halo's virial radius and therefore $M_{200} \equiv
800\pi \rho_c r_{200}^3/3$ as its virial mass.)

We generate our initial conditions using the methods described by
\citet{hern93}.  The initial positions of the particles in a halo are
drawn randomly from its density profile.  All of our initial haloes are
chosen to be $10^{12} M_{\odot}$.  In the case of the NFW profile the
total mass is formally infinite, so it is necessary to introduce a
truncation radius for the halo.  We choose to truncate our haloes at
$r=r_{200}$.  For our choice of $c=10$, the virial and truncation
radius is 162.6 $h^{-1}$ kpc and the scale radius is $r_s =r_{200}/c
=16.26 $ $h^{-1}$ kpc.  For the cored profile, we choose $r_{\rm core}
= 8 $ $h^{-1}$ kpc, $r_{\rm cut}=81$ $h^{-1}$ kpc, and truncate the
halo at $2 r_{\rm cut}$.  This choice of parameters allows us to start
with two types of galaxies, equal in mass and radius, but with very
different inner density profiles (core vs. cusp).

For the initial velocities of particles in the frame of a given halo,
we first use the Jeans equation (the first moment of the
collisionless Boltzmann equation) to relate the density profile
$\rho(r)$ and potential $\Phi(r)$ to the velocity dispersions
\citep{bt87}:
\begin{equation}
    \frac{d(\rho \sigma_r^2)}{dr}+2\frac{\beta}{r}\rho \sigma_r^2
    =-\rho \frac{d\Phi}{dr} \,,
\label{jeans}
\end{equation}
where the velocity anisotropy is parameterized by
\begin{equation}
   \beta=\beta(r) \equiv 1- \frac{\sigma_\theta^2 +\sigma_\phi^2}
   {2\sigma_r^2} \ ,
\label{beta}
\end{equation}
and $\sigma^2_i$ is the dispersion of velocity component $i$.  If we
further assume initial velocity isotropy, $\beta(r)=0$, we can
integrate Eqn.~(\ref{jeans}) to yield the dispersion of the radial
velocity as a function of radius,
\begin{equation}
    \sigma_r^2(r)=\frac{1}{\rho(r)} \int_r^{\infty} dR\, \rho(R)
    \frac{d\Phi}{dR} \,.
\label{veldisp}
\end{equation}
From there, we randomly draw individual velocity components from a
Gaussian distribution with variance $\sigma_r^2(r)$.  Following
\citet{hern93}, we impose the additional constraint $v \le 0.95
v_{\mathrm{esc}}$, which insures that all particles are initially
bound to the halo. 

We set up the two colliding haloes to be just touching at the
truncation radii initially.  The initial relative speed of the two
haloes is taken to be the speed one of the haloes would have if the
two were in circular orbit about their common centre of mass, i.e.,
$v_{\mathrm{rel}} = 2 \sqrt{G\mu/r_{\mathrm{rel}}}$, where $\mu$ is
the reduced mass and $r_{\rm rel}$ is the distance between the halo
centres.  If the two haloes were point particles, this configuration
would give $v_{\mathrm{rel}} =v_{\mathrm{esc}}/ \sqrt{2}$; all our
simulations are therefore relatively high-speed encounters for bound
orbits.  In this set-up, and with our choice of halo parameters, the
two haloes first cross centres of mass at approximately 1.2 Gyr, then
again at 2.2 Gyr.  This leaves ample time after the first encounter
for the haloes to interact and evolve within the 5.0 Gyr run.

Most of our simulations are head-on collisions with an impact
parameter $b=0$.  We have also performed two simulations in which the
two halo centres are initially offset by $b=0.5 r_{200}$ in order to
test the robustness of our conclusions. In terms of the dimensionless
parameter   
\begin{equation}
	\lambda=\frac{J|E|^{1/2}}{GM^{5/2}}	
\end{equation}
for the total mass $M$, energy $E$, and orbital angular momentum $J$,
this corresponds to $\lambda=0.042$.  In terms of the parameter
$\epsilon$, defined as the ratio of the angular momentum to that
needed for circular orbit at the same velocity and separation, this
set-up yields $\epsilon= 0.24$.  As a final test of the generality of
our findings, we performed two simulations of collisions involving
unequal mass haloes with cores. 

\begin{table}
\caption{\label{table-runs}
Properties of ten simulations of halo mergers.  All simulations
involve two $10^{12} M_{\odot}$ haloes (except where noted),
so $M_{200}=2 \times 10^{12} (N_{200} / N_p) M_{\odot}$}
\begin{tabular}{lllll}
\hline
\hline
\textbf{Name} &\textbf{b}$^1$
&$\mathbf{N_p}^2$
&$\mathbf{N_{200}}^3$ &$\mathbf{r(t_r)}^4$\\
 &($r_{200}$) & & &($h^{-1}$ kpc)\\
\hline
Core1 &0   &$2\times 10^5$ &$1.58\times 10^5$ &4.035\\
Core2 &0   &$2\times 10^6$ &$1.58\times 10^6$ &1.850\\
Core3 &0.5 &$2\times 10^6$ &$1.52\times 10^6$ &1.886\\
Core4$^5$ &0   &$1.33\times 10^6$ &$1.16\times 10^6$ &1.920\\
Core5$^5$ &0   &$1.33\times 10^6$ &$1.09\times 10^6$ &2.025\\
Cusp1 &0   &$2\times 10^5$ &$1.34\times 10^5$ &3.595\\
Cusp2 &0   &$2\times 10^6$ &$1.35\times 10^6$ &1.436\\
Cusp3 &0   &$1\times 10^7$ &$6.72\times 10^6$ &0.765\\
Cusp4 &0.5 &$2\times 10^6$ &$1.31\times 10^6$ &1.549\\
Mixed &0   &$2\times 10^6$ &$1.48\times 10^6$ &1.488\\
\hline
\end{tabular}
$^1$impact parameter, in units of $r_{200}$ for an initial halo\\
$^2$total number of initial particles\\
$^3$total number of particles within virial radius of remnant\\
$^4$radius above which the two-body relaxation time exceeds 5 Gyr\\
$^5$unequal mass ($M_1=3M_2$) collision; $M_1=10^{12} M_{\odot}$\\
\end{table}

\subsection{Simulation Parameters and Numerical Effects}

All of our simulations are done using {\tt GADGET}, a publicly available
N-body tree code \citep{gadget}.  {\tt GADGET} is available in serial and
massively parallel forms; we have tested both versions on a local linux
cluster  and the parallel version on the IBM SP2 computers at the
National Energy Research Scientific Computing Center (NERSC).
In these test runs, we use either $10^5$ or $10^6$ particles and run
for 5.0 Gyr. 
An important objective of the test runs is to check the level of
stability against two-body relaxation for an isolated halo
initially in equilibrium.  Two-body relaxation is a numerical
artifact of running simulations where each simulation particle is on
the order of $10^{60}-10^{75}$ times more massive than a typical
elementary particle dark matter candidate.  This leads to significant
particle encounters that one does not expect to find in a realistic
dark matter halo.  As a result, the two-body relaxation timescale
effectively puts a limit on the region of the halo that faithfully
represents the phase space distribution in the collisionless limit.

Though there has been some disagreement about what regions
can be considered reliable beyond the level of two-body relaxation
(e.g. \citealt{moore98, klypin01, fm01, power, diemand}), a conservative
criterion is that the local two-body relaxation timescale $t_r$, which
we define in terms of the circular velocity $V_{\mathrm{circ}}
=V_{\mathrm{circ}}(r)$, period $T(r)=2\pi r/V_{\mathrm{circ}}$, and
number of particles N (or equivalently mass M) interior to a radius r
as
\begin{eqnarray}
    t_r(r) &=& T(r_{200}) \frac{N}{8\ln{N}} \frac{r}{r_{200}}
    \frac{V_{\mathrm{circ}}}{V_{\mathrm{circ}}(r_{200})}\nonumber\\
           &=& \frac{\pi}{4} \frac{N}{\ln{N}}
    \sqrt{\frac{r^3} {G M(r)}},\label{twobody}
\end{eqnarray}
be at least as large as the age of the universe \citep{fm01,power}.
Since our simulations are purely gravitational, with no cosmological
expansion, the current age of the universe has no direct meaning for
the simulations themselves.  As a result, we use the slightly different
convergence criterion that the two-body relaxation time be longer
than the length of time the simulation has run.
We can invert Eqn.~\ref{twobody} to find $r(t_{r})$, the minimum radius
which satisfies $t_r > t_f$ for the final time of the simulation
$t_f$; we refer to this as our minimum converged radius (see
Table~\ref{table-runs}).

Our test runs of single isolated haloes show that the density profiles
are stable at all radii $r(t_r)<r \la 0.8 r_{200}$ for the full 5.0
Gyr of simulation: any change in the profile is under 1\%.  (Near
$r_{200}$ there are small deviations due to truncating the halo.)
Inside $r(t_{r})$, the behavior is dictated by the structure of the
halo.  NFW haloes have ``temperature inversions'' where the maximum
velocity dispersion is reached not at the centre but at a finite
radius.  As a result, two-body relaxation does not cause a contraction
of the central regions as is the case with the singular isothermal
sphere (the so-called gravothermal catastrophe \citep{bt87}) but
rather an expansion as energy is transfered from regions with higher
velocity dispersion to lower.  Any model with a cusp shallower than
$\gamma=2$ starts with a temperature inversion \citep{quinlan} and
therefore undergoes an expansion of the central region and a
concurrent reduction in central density.  The central 
expansion is then followed by a collapsing stage, initiated once the
temperature inversion is eliminated \citep{quinlan, hayashi03}.  This
process occurs on the evaporation timescale, which is much longer than
the length of our simulations at all radii.  We do see evolution as a
result of two-body interactions well within $r(t_r)$ for our isolated
haloes with $10^5$ particles, but seems to be minimal even at $0.5
r(t_r)$.  We conclude that our simulations will be converged on all
scales greater than $r(t_r)$ and are likely reliable to even smaller
radii.

After obtaining stable isolated haloes, we performed production runs
(all using the massively parallized version of {\tt GADGET}) of halo
collisions with different particle numbers to check for convergence in
the inner structure of merger remnants.  We did production runs with
zero impact parameter using $2\times 10^5$ and $2\times 10^6$ total
particles for the core-core collisions, and $10^7$ particles (in
addition to both lower particle number simulations) for an cusp-cusp
collision; for a core-cusp collision, we did one run with $2\times
10^6$ total particles.  For the two runs with non-zero impact
parameter, we used $2\times 10^6$ total particles.  Our two unequal
mass mergers were both core-core collisions with mass ratios of
$3:1$.  Both runs used the parameters described in Section 2.1 for the
more massive galaxy.  In one run (core4), we reduced all
characteristic radii by a factor of $3^{1/3}$, yielding a core radius
of $5.55$ kpc and a cut-off radius of $55.5$ kpc.  In the other run 
(core5), the less massive galaxy had the same length scales as the
more massive halo: $\mathrm{r_{core}} = 8$ kpc and $\mathrm{r_{cut}}=
81$ kpc, giving a halo that is spatially identical to the heavier one
but with the density uniformly lower by a factor of 3.  By using these
parameters, we can identify how a merging halo having either a
smaller core or a lower central density affects the merger process.

For all runs, we use a force softening of $0.175$ $h^{-1}$
kpc for our particles, the new cell-opening accuracy parameter
(errTolForceAcc) of 0.001, and timesteps that are inversely
proportional to the acceleration (timestep criterion 1 of {\tt
GADGET}).  (Our choice of errTolForceAcc is somewhat smaller than the
maximum allowable value suggested by \citet{power} in their recent
thorough investigation of numerical parameters in cosmological
simulations, but our tests have shown this value is necessary to
obtain results that are independent of machine architechture and
whether or not the run is parallel.)  A typical run involved between
7000 (core-core runs) and 10000 (cusp-cusp runs) timesteps.  Since the
crossing time $t_c= \sqrt{r_{200}^3/GM}$ is approximately 1 Gyr for
our haloes, an average timestep corresponds to approximately $5.9
\times 10^{-4}$ $t_{\mathrm{c}}$.  The parameters for all of the
production runs, as well as some results, are summarized in
Table~\ref{table-runs}.

\begin{figure}
\includegraphics[scale=0.5]{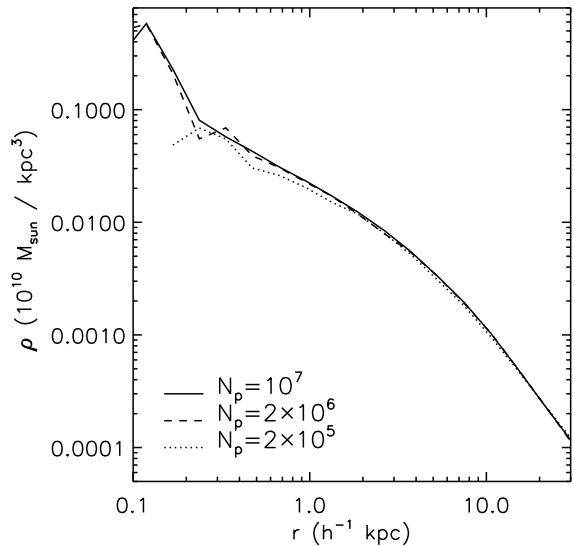}
\caption{\label{nfwres}Density profiles for 3 different resolution
simulations of the merger product of two cuspy haloes at 5 Gyr.  The
profiles are indistinguishable at large radii, but the profile for the
simulation with $2\times 10^5$ particles (dotted line) shows signs of
numerical relaxation within $\sim 2$ $h^{-1}$ kpc.}
\end{figure}

\begin{figure*}
\includegraphics{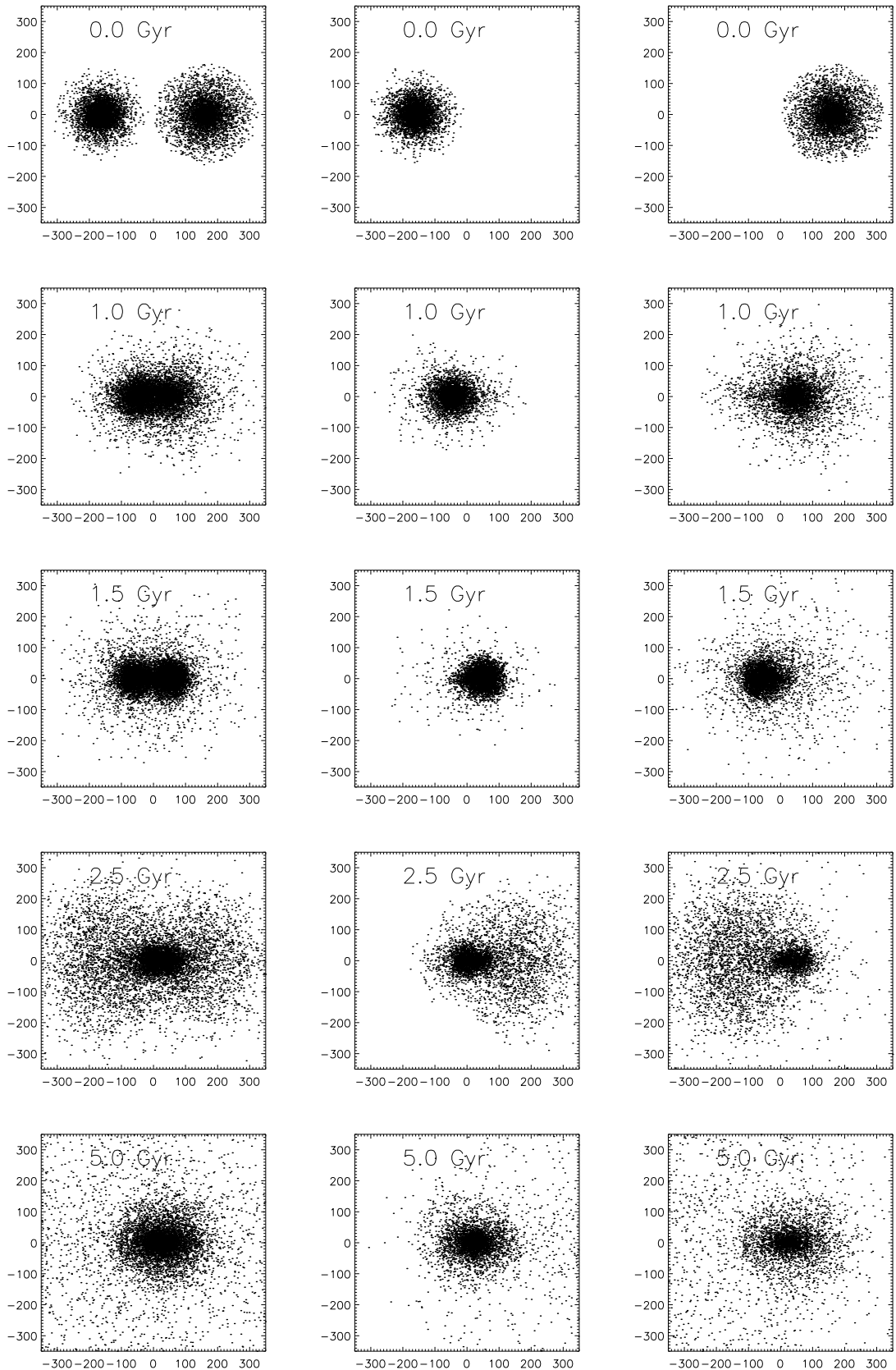}
\caption{\label{scatter}Scatter plot of five snapshots for simulation
``Mixed'' of a cored and a cuspy halo.  The collision is along
the horizontal axis, and the units of both axes are $h^{-1}$ kpc.
Left: both haloes. Middle: cored halo only.  Right: cuspy halo only.
The halo centres cross at $\sim 1.2$ Gyr, then recross at $\sim 2.2$
Gyr, at which point the relative centre of mass motion is almost
completely damped.} 
\end{figure*}

In all of our simulations, at least 65\% of the total initial mass
ends up within the virial radius of the final halo, meaning there are
over $6\times 10^6$ particles within $r_{200}$ in our highest
resolution simulation, a large enough number to reliably resolve
density contrasts exceeding $10^6$.  In all runs, at least 97\% of the
particles stays bound to the final halo.  All mergers appear to be
mostly completed by 3.5 Gyr, with the virial mass of the haloes
increasing by only $\sim 10$\% between 3.5 to 5.0 Gyr.
Fig.~\ref{nfwres} shows the density profiles at the final timestep for
the three cusp-cusp collisions.  The profiles of the remnants are
extremely similar at all radii greater than $2$ $h^{-1}$ kpc, and runs
``Cusp2'' and ``Cusp3'' have indistinguishable density profiles for
radii greater than $0.7$ $h^{-1}$ kpc.  These results indicate
two-body relaxation has not significantly affected any of our haloes at
radii larger than $2$ $h^{-1}$ kpc and that our two highest resolution
cusp-cusp collisions yield density profiles that are reliable on
sub-kiloparsec scales.

\subsection{Definition of Halo Centres}

The choice of how to define a halo's centre is crucial in analyzing
the final state of the merger remnant since quantities such as the
spherically averaged density profile and radial velocity dispersion
depend on the location of the centre.  Previous studies have mainly
used two methods for defining the centre of a halo: the most bound
particle(s) (MBP) and the centre of mass (COM).  For example, NFW
\citep{nfw96} use the COM; Moore et al. (1999a) use the MBP and
note that it gives very similar results to the COM; Pearce et
al. (1993) use the MBP and find that using the COM gives a
consistently shallower inner profile.

We adopt an iterative version of the COM technique, which we find to
be largely independent of initial assumptions.  We have tested this
method of computing a halo's centre against using either the MBP or
the COM of the five hundred most bound particles.  The resulting
spherically averaged density profiles are nearly identical at all
radii that we consider to be converged -- we find no evidence that
one method systematically yields steeper cusps than the other.  It is
also worth noting that in all cases, the density profiles using
different criteria for convergence of the centre of mass calculation
differ only at radii that we already consider to be unreliable due to
two-body relaxation and Poisson noise.  As noted earlier, the density
profiles are quite similar over a factor of 50 in particle number (see
Fig.~\ref{nfwres}).

\section{Results}

\subsection{Scatter Plots; Energy}

\begin{figure}
\includegraphics[scale=0.5]{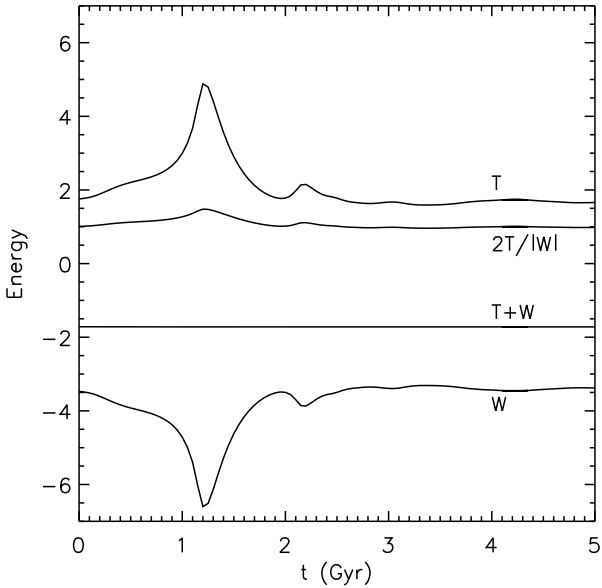}
\caption{\label{en_evol}Time evolution of the kinetic ($T$), potential
($W$), and total ($T+W$) energies, as well as the virial ratio
($2T/|W|$) for the cusp-core merger simulation.  The virial ratio
returns to unity soon after the first two encounters at $\sim 1.2$
and 2.2 Gyr.  The total energy is conserved to better than 0.4\%
during the run.  The units on the energy axis are arbitrary.}
\end{figure}

Fig.~\ref{scatter} shows five snapshots at $t=0$, 1, 1.5, 2.5, and 5
Gyr from the merger simulation of a cored and a cuspy halo (run
``Mixed'').  (The general sequence of events is the same for all three
types of mergers listed in Table~1.)  To illustrate the evolution of
the individual haloes, we also plot separately the cored halo (middle
panels) and cuspy halo (right panels).  During the encounter between 1
and 1.5 Gyr, the outer regions of the haloes pass through each other
mostly unimpeded; the main effect in the outer portions is for the
halo to become more extended.  The panels at 2.5 Gyr show that the
cuspy halo's outer region is more easily stripped away than that of
the cored profile; this is as expected, since the NFW profile is more
dense in the centre but less tightly bound towards the outer region.

The merger remnants of the head-on collisions are noticeably prolate
(extended along the collision axis) with axis ratios of $\sim$ 1.6:1:1
at half-mass radius.  This result is quite similar regardless of the
initial density profile.  The two offset simulations ``Core3'' and
``Cusp4'' with orbital angular momentum result in mildly triaxial
remnants.

Fig.~\ref{en_evol} shows the evolution of the kinetic, potential and
total energies as well as the virial ratio for the same run in
Fig.~\ref{scatter}.  The initial crossing of halo centres is seen to
occur at $\sim$ 1.2 Gyr, followed by a second crossing at $\sim$ 2.2
Gyr.  Subsequent oscillations are strongly damped, and the system
moves towards a steady state in virial equilibrium.  The major
features in Fig.~\ref{en_evol} are common to all production runs
listed in Table~1.  All runs conserve energy to better than 0.4\% and
have virial ratios $2T/|W|$ within 2\% of 1.0 by the final timestep.

\subsection{Density Profile}

\begin{figure*}
\includegraphics{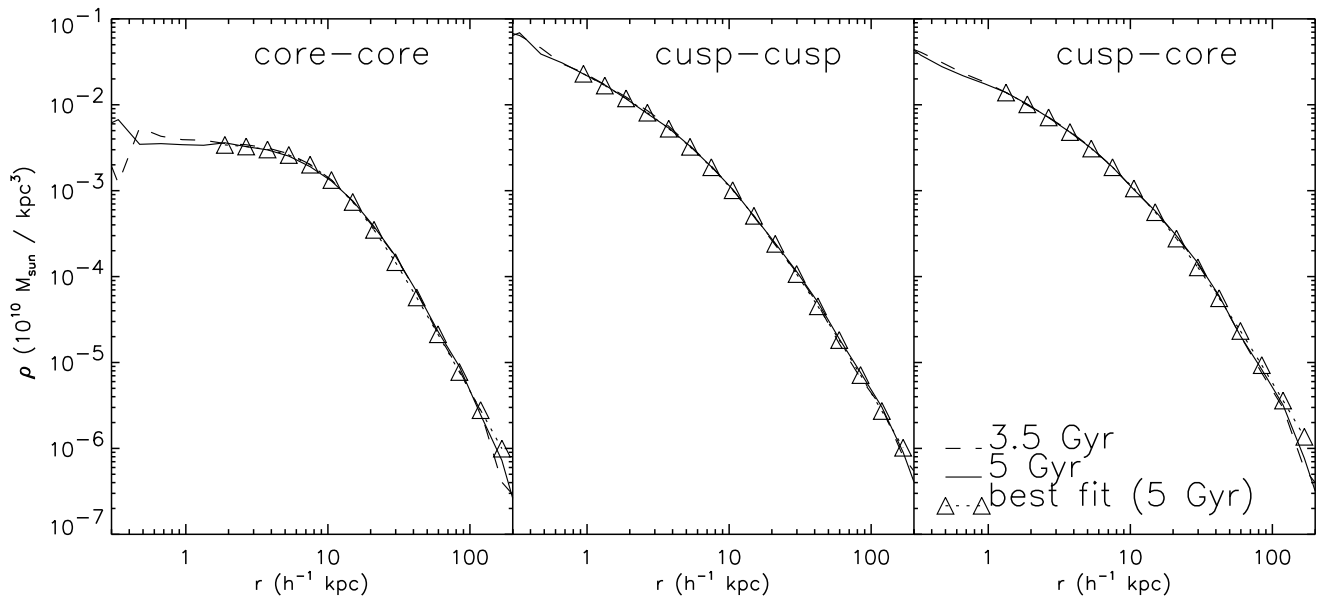}
\caption{\label{densmulti}Density profiles for the core-core (left),
cusp-cusp (centre), and cusp-core (right) simulations with $2 \times
10^6$ total particles, along with fits from
Eqns.~\ref{eqncore} and \ref{eqnnfw}.  Both the cusp-cusp and the
cusp-core collisions are well fit by an NFW-like profile with an inner
cusp of $\gamma \approx 0.7$}
\end{figure*}

\begin{figure}
\includegraphics[scale=0.5]{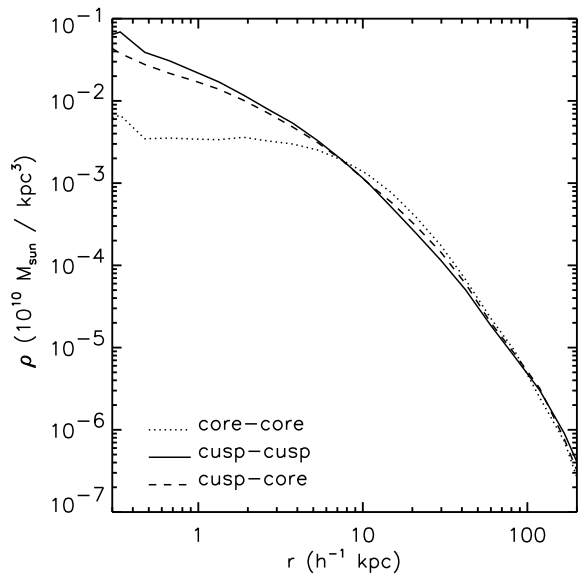}
\caption{\label{densall}Density profiles for the core-core (dotted),
cusp-cusp (solid), and cusp-core (dashed) runs with $2\times
10^6$ particles at 5.0 Gyr.  The cusp-core collision results in a
remnant with a density profile much more similar to the cusp-cusp
collision than to the core-core one.}
\end{figure}

In Fig.~\ref{densmulti}, we plot the spherically-averaged density
profiles $\rho(r)$ of the merger products at 3.5 Gyr (dashed curve)
and 5.0 Gyr (solid curve) for each of the three types of collisions:
core-core (left), cusp-cusp (middle), and core-cusp (right).  The
density profiles in all cases are quite stable and constant from 3.5
to 5.0 Gyr, indicating that the merging process is essentially
complete by 3.5 Gyr (also evident in Fig.~\ref{en_evol}) and that our
simulations with $2\times 10^6$ particles do not suffer much from
relaxation effects for the radii shown in Fig.~\ref{densmulti}.  The
dotted curves in Fig.~\ref{densmulti} show our fitting results to be
discussed below.  For comparison, we overlay in a separate plot
(Fig.~\ref{densall}) the final profiles of the merged haloes at 5 Gyr
from the three collisions.

\begin{table}
\caption{\label{table-fits}Fits (to Eqn.~\ref{eqnnfw}) for
three mergers involving cuspy haloes at 5 Gyr.} 
\begin{tabular}{llll}
\hline
\textbf{Name} &\textbf{A} &\textbf{B} &\textbf{C}\\
& ($10^{10} M_{\odot} \mathrm{kpc}^{-3}$) & ($h^{-1}$ kpc) & (cusp slope)\\
\hline
Cusp2 &0.00593 &9.75 &0.672\\
Cusp3 &0.00655 &10.0 &0.696\\
Mixed &0.00425 &12.1 &0.647\\
\hline
\end{tabular}
\end{table}

The final product of the core-core collision does {\it not} show signs
of moving towards an NFW profile: the remnant has a core of
approximately the same size as in the original haloes.  In order to
quantify our results, we have tried to fit the final profile to the
same functional form as used for the initial cored profile
(Eqn.~\ref{hernquist}).  We find the inner region of the remnants well
fit by a 
core radius of $r_{\rm core}=7.8$ $h^{-1}$ kpc (for run ``Core2''),
which is nearly identical to the initial $r_{\rm core}=8$ $h^{-1}$
kpc, but the density in the outer region of the remnants falls off
more gradually, as $r^{-3}$ instead of the initial exponential cutoff
in Eqn.~(\ref{hernquist}).  Overall, we find the merged halo of
core-core collisions better fit by 
\begin{equation}
   \rho(r)= \frac{A}{(r^2+B^2)^{3/2}} \,,
\label{eqncore}
\end{equation}
with $A\approx 5 \times 10^{10} M_{\odot}$ and $B\approx 11$ $h^{-1}$
kpc, than by the original profile (\ref{hernquist}).  Since the
remnants have virial radii about 20\% larger than 
their progenitors, this means the relative size of the core in the
final product is smaller, a result consistent with those of earlier
simulations \citep{white78, pearce93}.  The high spatial and mass
resolution used here allows us to simulate scales more than five times
below the core radius and to conclude that the core is preserved
during one head-on major merger.

In the cusp-cusp collision, the product is essentially a new cuspy
halo with an NFW-like form, which we find well fit by 
\begin{equation}
    \rho(r)=A \frac{B^3}{r^C(B+r)^{3-C}}
\label{eqnnfw}
\end{equation}
for radius down to the relaxation scale $r=0.7$ $h^{-1}$ kpc for the
$N_p=10^7$ particle run.  However, the best-fitting value for the
logarithmic slope of the cusp is shallower than 1, with $C\approx
0.7$, perhaps hinting that collisions of galaxies serve to reduce the
cusp at very small radii.  The best-fitting parameters for the density
profiles of our two highest resolution cusp-cusp collisions, as well
as for the core-cusp collision, are summarized in Table~\ref{table-fits}.

In the case of the core-cusp merger, any hint of the core is entirely
eliminated.  The remnant has an overall density profile much closer to
NFW than to the original core model and again well fit by
Eqn.~(\ref{eqnnfw}) with an inner slope of $C\approx 0.65$.

\begin{figure*}
\includegraphics{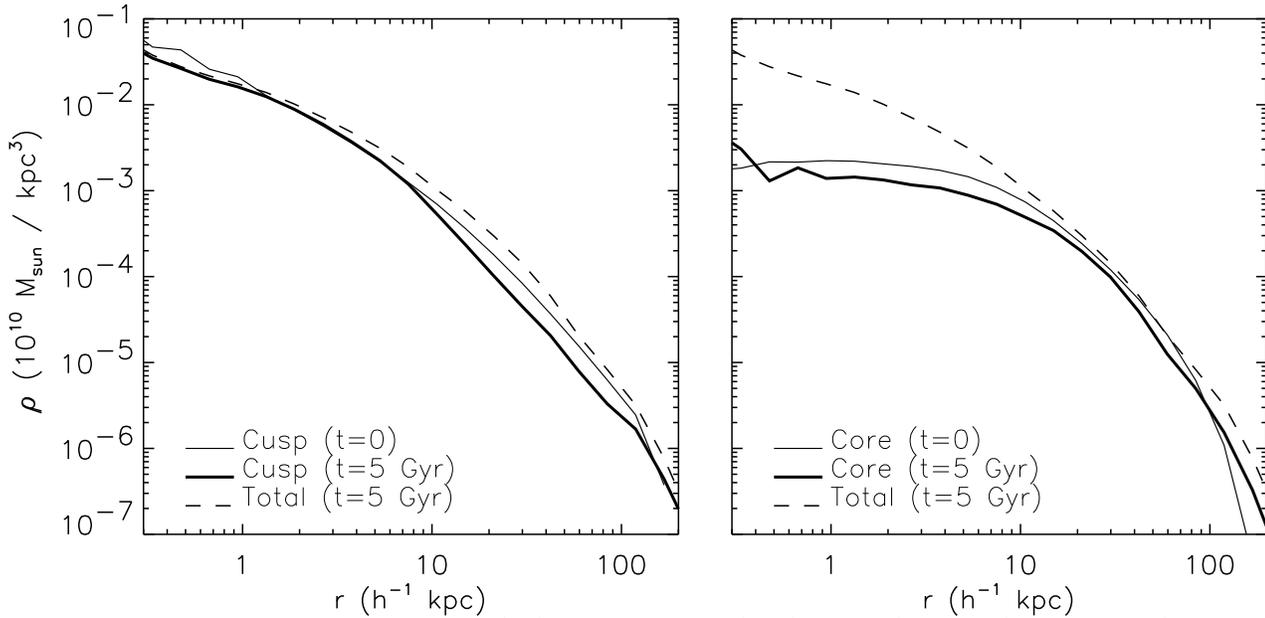}
\caption{\label{rhomix}Density profiles of the cuspy particles (left)
and core particles (right) initially (think solid) and at 5 Gyr (thick
solid) in the core-cusp collision simulation.  The particles
originating from the cuspy (cored) halo clearly maintain the cuspy
(cored) profile after the merger.  The dashed line shows the total
density profile of the merged halo at 5 Gyr.}
\end{figure*}

To understand the core-cusp merger in more detail, we calculate the
spherically-averaged density profile for particles in each halo
separately, first at $t=0.0$ Gyr and again at $t=5$ Gyr; the results
are shown in Fig.~\ref{rhomix}.  Interestingly, the particles
originating from the cuspy halo maintain a cuspy, NFW-like profile
(left panel), while the particles originating from the cored halo
retain their original density profile to a remarkable extent as well
(right panel).  The final density profile is dominated by the cusp
particles in the inner $\sim 10$ kpc and by the core particles from
$\sim 10-100$ kpc; outside of $\sim 200$ kpc, both types of particles
seem to contribute approximately equally (also see Fig.~\ref{track2}).

\begin{figure}
\includegraphics[scale=0.53]{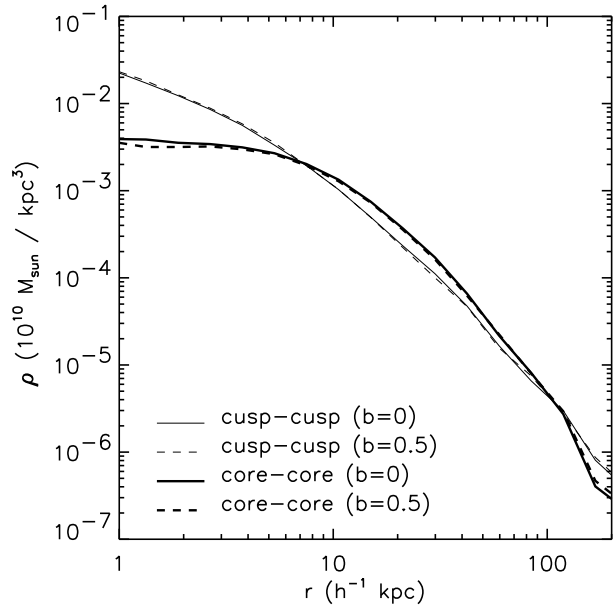}
\caption{\label{rho_ang}Density profiles for cusp-cusp (thin lines)
and core-core (thick lines) runs with $N_p=2\times 10^6$ for $b=0$
(solid) and $b=0.5 r_{200}$ (dashed) at 3.5 Gyr.  The profiles are
extremely similar, indicating that introducing a moderate amount of
orbital angular momentum does not significantly impact the merger
remnant.}
\end{figure}

The results above are all for head-on collisions with a zero impact
parameter.  In Fig.~\ref{rho_ang} we compare the density profiles of
the non-zero impact parameter runs with the corresponding zero impact
parameter simulations.  The profiles are almost exactly the same,
leading us to the conclusion that introducing moderate amounts of
orbital angular momentum into a merging system will not influence our
results.

\begin{figure}
\includegraphics[scale=0.53]{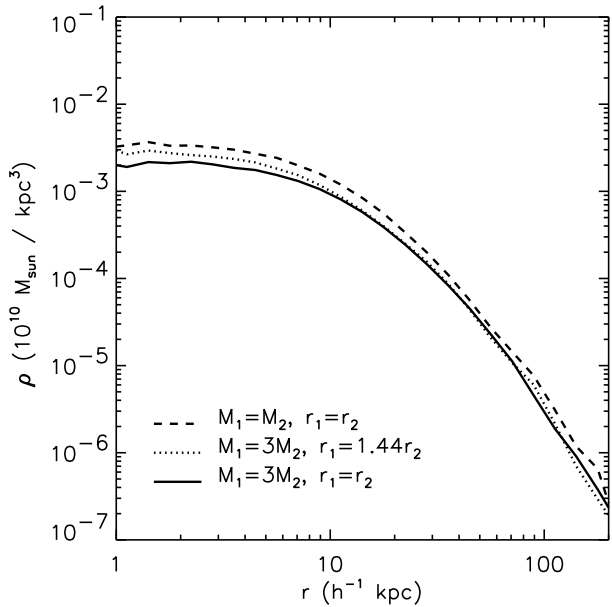}
\caption{\label{rho_unequal}Density profiles for three different
core-core mergers at $3.0$ Gyr.  Dashed line: run Core2 (equal mass,
both core radii $=8$ kpc).  Dotted line: run Core4 ($M_1=3M_2$ with
core radii $r_1=8$ kpc, $r_2=5.55$ kpc).  Dashed line: run Core5
($M_1=3M_2$ with core radii $r_1=r_2=8$ kpc).  All three mergers
produce similar remnants with cores that are unreduced relative to the
larger of the two initial cores, showing that mergers of cored haloes
result in cored remnants.}
\end{figure}

To ensure that the results are not specific to equal-mass mergers, we
plot the density profiles of our two $3:1$ mass mergers in
Fig.~\ref{rho_unequal}, along with the density profile of the remnant
in an equal mass merger for reference.  The remnants of both unequal
mass mergers retain cores of $\sim 8$kpc, the same result we found in
the equal mass case.  In fact, the final density profiles in all
three are remarkably similar; this shows that core-core major mergers
produce cored remnants.

\subsection{Structure of Central Region}

Although our simulations are collisionless, resulting in a
phase space density that must be conserved due to Liouville's theorem,
we have no \textit{a priori} reason to believe the configuration and
momentum space distributions will be independently conserved,
i.e. that a merger of two cored (cuspy) haloes will necessarily result
in a cored (cuspy) remnant.  A system's coarse-grained distribution
function $\overline{f}$ is constrained to be non-increasing as a
function of time \citep{bt87}, and thus $\overline{f}$ in the final
state is required at every point to be less than the maximal initial
$\overline{f}$, but this requirement does not prevent cusps from
becoming cores or vice-versa in collisionless simulations.  In fact,
violent collisions provide strongly fluctuating potentials that cause
particles to rapidly exchange energy with the background, a process
that could quickly affect the phase space structure of both initial
haloes. Since the main objective of our study is to investigate the
effects of major mergers on the phase-space distributions of dark
matter in the inner regions of galaxy haloes, we examine this region
in more detail in this section.

In the core-core and cusp-cusp collisions, we find that as expected,
both progenitors contribute essentially equally to the particles at
each radius of the merged halo.  As Figs.~\ref{densmulti} and
\ref{densall} show, the inner density profile of the merger remnants
is similar to their progenitors, i.e., cored haloes remain cored; cuspy
haloes remain cuspy.

\begin{figure}
\includegraphics[scale=0.55]{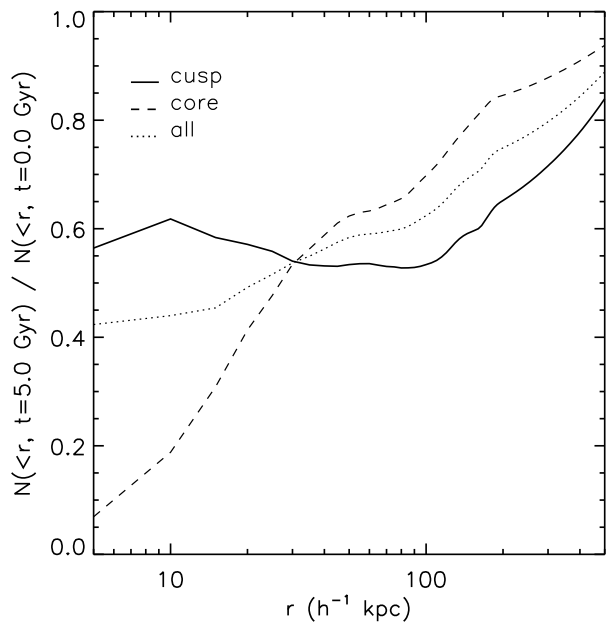}
\caption{\label{track1}Fraction of particles starting out within
radius $r$ at t=0 Gyr that also end up within radius $r$ at t=5 Gyr
(from run Mixed).  The cuspy particles (solid line) are much more
likely to stay within the inner 10 kpc than the cored particles
(dashed line).}
\end{figure}

The situation is more intriguing for the mixed core-cusp collision.
In order to elucidate the inner structure of this type of mergers, we
first examine where the initial inner particles in each halo end up in
the merged halo after 5 Gyr.  (The results at 3.5 Gyr are very
similar.)  Fig.~\ref{track1} shows the fraction of particles in each
halo that start out within radius $r$ initially and also remain within
$r$ at 5 Gyr.  (By construction, this fraction approaches 1 at large
$r$.) It shows that in the innermost several kiloparsecs of the merged
halo, about 60\% of the cusp particles (solid curve) that originated
from within a given radius stay within that same radius, while very
few of the core particles (dashed curve) remain.  We believe it is
this ``lack of mobility'' of the cusp particles that gives the cuspy
halo a stronger influence at the final remnant's centre and forces the
final density profile to be cuspy rather than cored.

\begin{figure}
\includegraphics[scale=0.55]{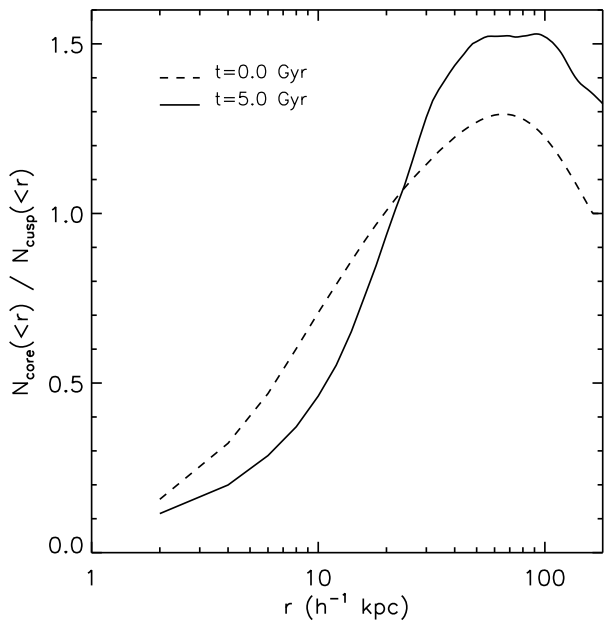}
\caption{\label{track2}Ratio of the number of
core particles within radius r to number of cusp particles within
radius r for the cusp-core collision at t=0 Gyr (dashed) and t=5 Gyr
(solid).  The remnant is even more centrally dominated by cusp
particles than would be the case if the two initial haloes were
superimposed, further showing how the core particles dominate the
inner structure of the remnant.}
\end{figure}

Next we examine the relative contribution of cusp vs core particles to
the total density (or potential) in the core-cusp collision.
Fig.~\ref{track2} shows the ratio of core to cusp particles initially
(dashed curve) and at 5.0 Gyr (solid curve).  The figure shows that
the cusp particles dominate the core particles number-wise in the
centre initially (if the two galaxies were superimposed with a common
centre), and that at 5 Gyr, the central region of the merger remnant
becomes even more dominated by cusp particles.  This result implies
that the tightly bound particles in a cusp tend to stay bound
together, while a core is much easier to disrupt.  This complements
the information from Fig.~\ref{track1} by showing that for all
particles of a given type, regardless of origin, the cuspy particles
tend to take over the central region.  Moving outward, the core
particles begin to contribute more to the remnant.  Note that the
radius where the number of core particles is equal to the number of
cusp particles stays almost constant at $r\approx 20 h^{-1}$
kpc.  This is one indication that there is not a major reconfiguration
of the particles occuring during the course of the simulation.  We
have also calculated the quantities in Figs.~\ref{track1} and
\ref{track2} at 3.5 Gyr and find them to be very similar to those
found at 5 Gyr. This corroborates the evidence in Fig.~\ref{en_evol}
that the merger is essentially completed by 3.5 Gyr.

\subsection{Velocity Profile}

\begin{figure*}
\includegraphics{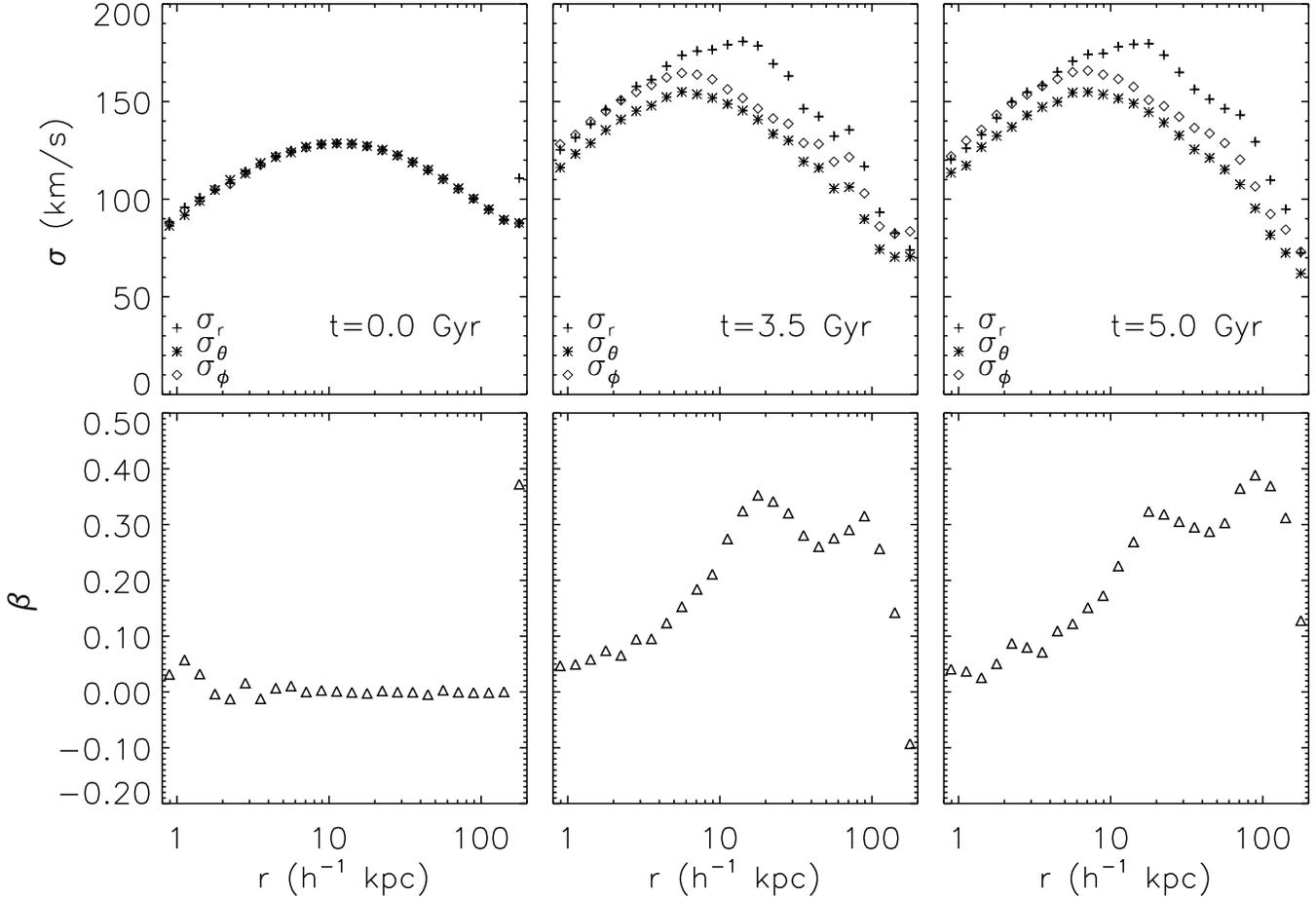}
\caption{\label{sigma}
Velocity dispersions $\sigma_r$, $\sigma_\theta$, $\sigma_\phi$ (top)
and velocity anisotropies (bottom), $\beta \equiv
1-(\sigma_{\theta}^2+\sigma_{\phi}^2)/2 \sigma_r^2$, as a function of
halo radius $r$ from the cusp-cusp merger run with $10^7$ particles
(run ``Cusp3'').  Three time outputs are shown: 0 Gyr (left), 3.5 Gyr
(middle), and 5 Gyr (right).  All three velocity dispersions increase
after the collision, and the initially isotropic state becomes
somewhat anisotropic outside the central region.}
\end{figure*}

The particle velocities in our initial haloes are isotropic by
construction, but those in our final haloes all exhibit some degree of
anisotropy.  For the cusp-cusp collision, Fig.~\ref{sigma} shows the
velocity dispersions (top panels) and velocity anisotropy (bottom
panels) as a function of halo radius $r$ at $t=0$, 3.5, and 5 Gyr for
the simulation with $10^7$ particles (run ``Cusp3'' in Table~1).
The velocities remain mostly isotropic (i.e. $\beta=0$) near the centre
of the merger remnant, but become mildly anisotropic ($\beta < 0.40$)
with increasing radius due to radial infalls.

We have also examined the velocity distribution function for the same
cusp-cusp run.
\begin{figure*}
\includegraphics{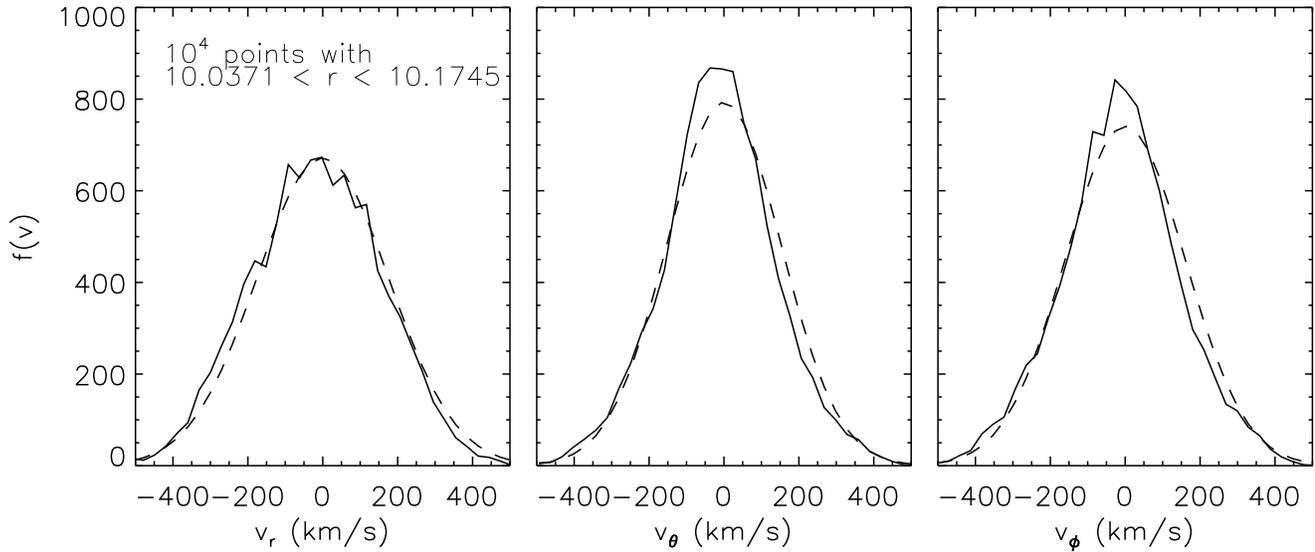}
\caption{\label{fv} Velocity distribution functions (solid) for $v_r$
(left), $v_\theta$ (middle), and $v_\phi$ (right) for $10^4$
particles in radial bin $10.0371 < r < 10.1745$ kpc from the
cusp-cusp run with a total of $10^7$ particles at 5 Gyr.  
The distributions are well-fit by Gaussians (dashed)
with width $(\sigma_r, \sigma_\theta,\sigma_\phi)=(177,153,161)$ km
s$^{-1}$ calculated from the particle velocities.}
\end{figure*}
We find that in the central region of the merged haloes, the
velocities are quite Gaussian in addition to being isotropic;
Fig.~\ref{fv} shows $f(v_i)$ vs. $v_i$ for $i=r, \theta, \phi$ for
$10^4$ particles in a radial bin arbitrarily centred on 10.106
kpc.  The distribution function of $v_r$ (left panel) is well fit by a
Gaussian (dashed line) with width $\sigma_r=177$ km s$^{-1}$, computed
from the particles in that bin. The distributions of $v_\theta$
(middle panel) and $v_\phi$ (right panel) are also reasonably well fit
by the corresponding Gaussians: $\sigma_{\theta} =153$ km s$^{-1}$ and
$\sigma_{\phi} =161$ km s$^{-1}$.  Moving away from the centre, we
find the radial distribution becomes substantially less Gaussian and
the angular distributions remain mostly Gaussian in the intermediate
portion of the halo.  Near the virial radius, however, none of the
distributions are Gaussian.

The core-core collision exhibits similar trends as in
Figs.~\ref{sigma} and \ref{fv}, so we will not show the results here.

\begin{figure*}
\includegraphics{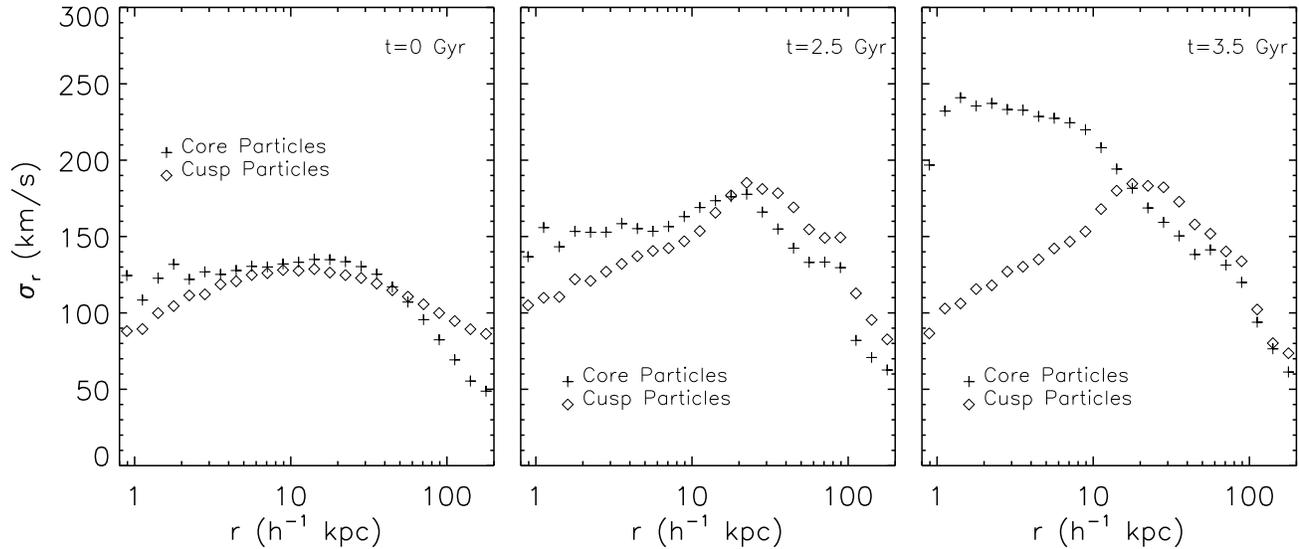}
\caption{\label{sigmam} Radial velocity dispersion $\sigma_r$ as a
function of halo radius $r$ from the core-cusp merger run. Three time
outputs are shown: 0 Gyr (left), 2.5 Gyr (middle), and 3.5 Gyr (right).
Particles originating from the cored (crosses) and cuspy haloes
(diamonds) are shown separately.  These two types of particles have
comparable $\sigma_r$ initially, but the core particles end up with
significantly higher $\sigma_r$ in the inner 10 kpc than the cusp
particles after the merger.  The cusp particles are also heated,
especially at $r\sim 10-30$ $h^{-1}$ kpc, as a result of the merger.}
\end{figure*}

For the core-cusp collision, we plot in Fig.~\ref{sigmam} the radial
velocity dispersion $\sigma_r$ as a function of radius $r$ at
$t=0,2.5$, and 3.5 Gyr for the core (crosses) and cusp (diamonds)
particles separately.  (Little evolution occurs after 3.5 Gyr.)  The
dispersion of the particles in the cuspy halo changes moderately
between 0.0 and 3.5 Gyr, with the peak in $\sigma_r$ rising and moving
outward but the overall shape maintained.  For the particles in the
cored halo, however, the velocity dispersion near the centre has
almost doubled to $\sigma_r \approx 230$ km s$^{-1}$ after the merger.

To understand why the velocity dispersion of the core particles
increases toward the centre of the merged halo upon merging with a
cuspy halo, it is instructive to consider a situation where the two
haloes are spherically symmetric and share a common centre.
In this case, the potential $\Phi$ is a function of the radial
coordinate only and can be written as $\Phi(r) =\Phi_1(r) + \Phi_2(r)$,
a sum of the individual potentials $\Phi_1(r) =\Phi_{\mathrm{core}}$
and $\Phi_2(r) =\Phi_{\mathrm{cusp}}$.  For particles in either the
cored or cuspy halo, we then have
\begin{eqnarray}
    \sigma_r^2(r) &=& \frac{1}{\rho(r)} \int_r^{\infty} dR\ \rho(R)
    \left( \frac{d\Phi_1} {dR} + \frac{d\Phi_2} {dR} \right) \nonumber\\
    &=& \frac{1}{\rho(r)} \int_r^{\infty} dR\, \frac{G\rho(R)} {R^2}
     M_{\mathrm{tot}}(R),
\end{eqnarray}
where $M_{\mathrm{tot}}(R)=M_1(R) +M_2(R)$ is the total mass interior
to radius $R$.  From the point of view of the core particles, adding
the cuspy particles has the effect of adding considerable extra mass
and thus deepening the potential.

\begin{figure*}
\includegraphics{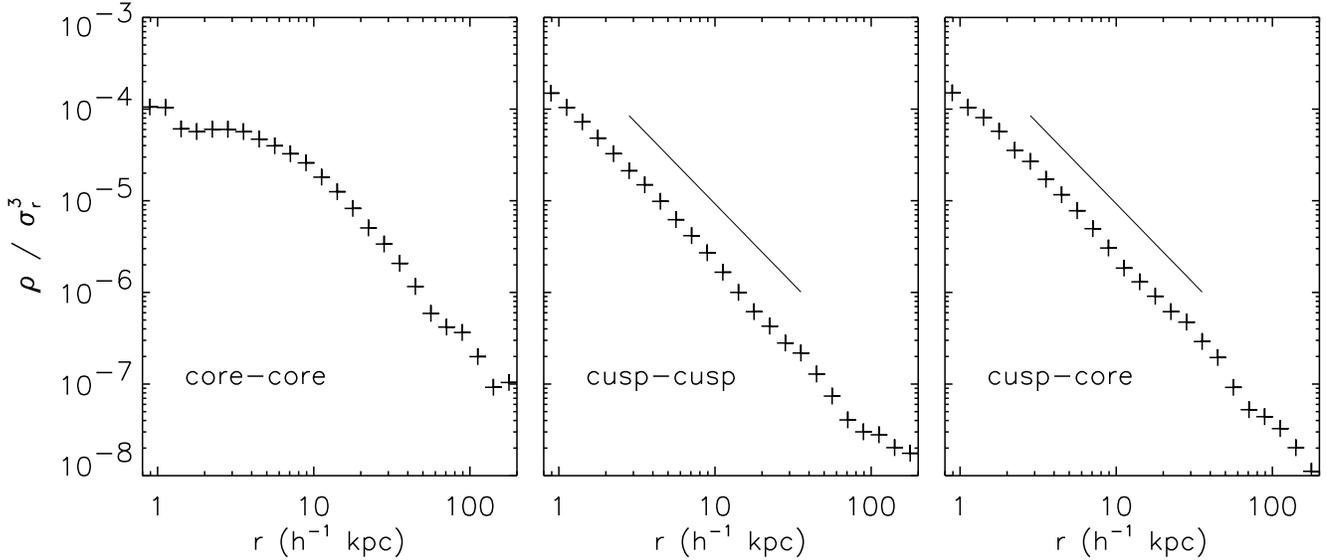}
\caption{\label{psd}Local phase space density, defined as $F=\rho(r) /
\sigma^3(r)$, for the $N_p=2 \times 10^6$ simulations of core-core
(left), cusp-cusp (centre), and cusp-core (right) mergers. The
densities are normalized to be equal to that of the cusp-cusp
simulation at $r=1$ $h^{-1}$ kpc.  The cusp-cusp and cusp-core
collisions give phase space densities that can be approximated by a
single power law, $F \propto r^{-1.75}$ (overplotted line), for $r \le
r_{200}$ while the core-core collision cannot.} 
\end{figure*}

If the density profile of the core particles remains roughly constant,
then adding extra mass in the form of cuspy particles essentially
heats the core particles to higher a velocity dispersion, as seen in
Fig.~\ref{sigmam}.  The reverse is also true: the core particles also
heat the cusp particles.  This effect is actually seen in
Fig.~\ref{sigmam}, as the cuspy particles' dispersion increases
relative to the original value, especially noticeably near the
original peak of $\sim 10$ $h^{-1}$ kpc.  Since the cuspy particles
dominate the total mass at radii less than this, however, the core
particles are a small perturbation and do not change the cuspy
particles' dispersion significantly.  The change in the peak of the
NFW particles' dispersion from $\sim 10$ to $\sim 30$ $h^{-1}$ kpc can
also be understood in this manner, since the core particles actually
contribute more of the interior mass at this radius, as shown in
Fig.~\ref{track2}.

\subsection{Phase Space Profiles}

The previous subsections discussed the density and velocity profiles
separately.  It is also instructive to examine the local phase space
density $\rho(r)/\sigma^3(r)$, which is shown in Fig.~\ref{psd} for
the three types of mergers studied in this paper.  We find that for
both the cusp-cusp (middle panel) and core-cusp (right panel) mergers,
although the individual $\rho(r)$ and $\sigma(r)$ are not a pure
power-law in $r$ (see Figs.~\ref{densmulti}, \ref{sigma}, and
\ref{sigmam}), 
the ratio is interestingly
well-approximated by a power law $\rho/\sigma^3 \propto r^{-\alpha}$,
where the best-fitting slope is $\alpha\approx 1.75$ for both mergers.
The core-core merger (left panel), on the other hand, shows a core in
the phase space density since both $\rho(r)$ and $\sigma(r)$ are
constant at $r<r_{\rm core}$.

Our power-law results can be compared with \citet{tn01}, who have
suggested that the phase space structure of haloes is intimately
related to that of self-similar collapse onto spherical density
perturbations in an expanding universe.  They solved the Jeans
equation under the assumption of isotropy and power-law $\rho /
\sigma^3 \propto r^{-\alpha}$ and found that haloes with the NFW
profile are well-approximated by the same power law, $\alpha= 1.875$,
as the self-similar solutions in secondary spherical infall models
\citep{bert85,fg84,gott75,gunn77}.  Since the phase space density is
inversely proportional to the local entropy, Taylor \& Navarro were
led to the conclusion that CDM halo structure may be driven toward
the most sharply peaked phase space distribution allowable for a
monotonically decreasing density profile.  For a power law phase space
density, this solution yields density profiles that are very similar
to the NFW profile but with a slightly shallower inner cusp,
$\rho\propto r^{-0.75}$ rather than $r^{-1}$.

Using the power-law result $\rho / \sigma^3 \propto r^{-1.75}$ in
Fig.~\ref{psd}, we obtain from the Jeans equation a density profile
$\rho\propto r^{-0.72}$, which is consistent with the best-fitting
inner slope shown in Sec.~3.2.  One can also look at the question in
reverse, i.e., given a power-law density $\rho\propto r^{-\gamma}$,
what solutions are admitted for the phase space density?  Calculations
show that $\gamma \approx 0.7$ cusps lead to phase space densities
well approximated by $\rho/\sigma^3 \propto r^{-1.75}$ as $r
\rightarrow 0$, but the slope of $\rho/\sigma^3$ is not strictly a
constant beyond the central region (Singh \& Ma, in preparation).

\section{Discussion}

Our simulations offer insight into the mechanisms that drive a merging
system toward an equilibrium structure, e.g., violent relaxation
\citep{lynden-bell} and phase mixing.  Violent relaxation is due to a
time-varying potential which allows the non-conservation of individual
particles' energies.  The relative motion of the two haloes at the
start of our simulations provides such a fluctuating potential. From
Fig.~\ref{en_evol}, it is clear our mergers involve a rapid transfer
of orbital energy to internal energy of the individual haloes (and
subsequently the merger remnants).  The decay of orbital motion occurs
approximately on the crossing timescale, with essentially all energy
in the form of the remnants' internal energy by 2.5 Gyr.  That the
orbital motion is so heavily damped and that the remnant haloes do not
expand or collapse significantly indicate that the time-varying
potential necessary for violent relaxation is eliminated relatively
quickly.  One possible exception is that density waves could enable
violent relaxation to persist even once the haloes have merged.  Phase
mixing tends to damp these waves strongly, however, with a decay
timescale of $\tau \approx \lambda / \sigma$ for a perturbation of
size $\lambda$.  Taking $\sigma_r=150$ km s$^{-1}$ to be typical for
our merged haloes, this means that phase mixing will damp out all
perturbations smaller than the virial radius within $\sim 1$ Gyr of
the haloes merging.
In our simulations, this seemingly short timescale for violent
relaxation is sufficient to drive the velocity distribution function
in the inner region of the merged halo to a Gaussian form, as
predicted by statistical reasoning (e.g. \citealt{nakamura}).  That the
distribution function becomes less Gaussian with increasing radius is
also in line with expectations because much less mixing is possible in
the outer regions of the halo on the timescales we are studying.

\begin{figure}
\includegraphics[scale=0.50]{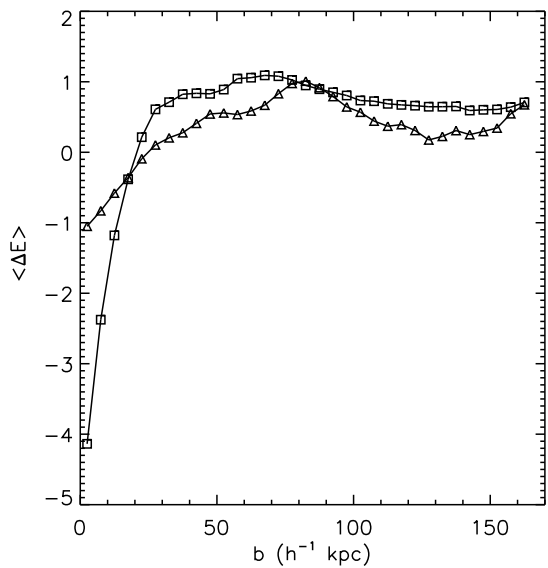}
\caption{\label{impact}Average energy change (in arbitrary units)
versus impact parameter for the cusp-core collision.  Squares are
particles from the cuspy halo, triangles are particles from the cored
halo.  All particles with small impact parameters tend to lose energy,
while those with large impact parameters gain energy.  The effect is
more dramatic for cuspy particles than cored ones.} 
\end{figure}

Following the energy distribution through the collision can help in
understanding the structure of the final product.  A particularly
interesting quantity is the change in energy $\Delta E$ for a particle
from initial to final state as a function of position.  \citet{funato}
describe an energy segregation in their collision simulations:
particles with higher (less negative) energy tend to gain more energy
than particles with lower energy.  This means a particle near the
centre of an initial halo should, on average, lose energy (i.e. become
more tightly bound) relative to a particle near the virial radius of a
halo.  We follow the energy change of each particle, and plot in
Fig.~\ref{impact} the change in energy as a function of impact
parameter for the core-cusp run ``Mixed''.  As Funato et al. describe,
the particles with small impact parameters tend to lose energy, while
the particles with large impact parameters tend to gain it.  The
effect is more dramatic for the cuspy particles than for the core
particles, perhaps because the inner portion of the cuspy halo
is initially more tightly bound but with an essentially equal
velocity dispersion profile.  As noted by \citet{mh03}, however, this
energy segregation does not prevent the system from equilibrating via
violent relaxation, at least in the central region.

\begin{figure}
\includegraphics[scale=0.48]{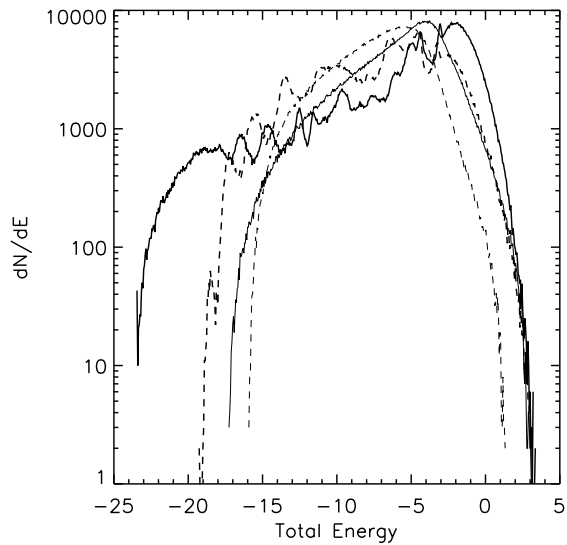}
\caption{\label{ehist}Differential energy distribution for the
cusp-core merger.  the solid lines show the cusp particles and the
dashed lines show the core particles; thin lines indicate the initial
distributions and thick lines the final distributions.  Both
distributions are broadened as a result of the merger, and shell-like
structures are visible in the final distributions.  The energy units
are arbitrary.}
\end{figure}

Another interesting quantity is the differential energy distribution
$dN/dE$, which can be written as a product of the phase-space
distribution function $f(E)$ and the
density of states $g(E)$ \citep{bt87}:
\begin{equation}
\frac{dN}{dE}=f(E)g(E) \,.
\end{equation}
The shape of $dN/dE$ is therefore \emph{not} the same as the shape of
$f(E)$.  In Fig.~\ref{ehist} we plot $dN/dE$ for the core-cusp merger,
for both types of particle, at the start of the simulation (thin
curves) and at 5 Gyr (thick curves).  It shows that $dN/dE$ is
broadened, a natural outcome of mergers where the range of energies
accessible to an individual particle can change drastically.  The
wave-like features in $dN/dE$ at 5 Gyr are indications of shells of
particles formed via phase-mixing \citep{sh92}.  We note that unlike
$f(E)$, which generally decreases with increasing $E$, the shape of
$dN/dE$ is skewed towards higher energies because of the density of
states.  Both $f(E)$ and $dN/dE$ have been studied analytically for
isolated NFW \citep{widrow00} and King haloes \citep{bt87} in
equilibrium.  The overall shape of $dN/dE$ resembles that shown in
Fig.~\ref{ehist}.

\section{Summary}

We have simulated major mergers of galaxy haloes with both cuspy and
cored inner density profiles using {\tt GADGET}, a versatile N-body tree code.
Our findings can be summarized as follows:

\begin{enumerate}

\item Mergers of two cored haloes of equal mass result in a cored halo.
The core radius of the merged halo is nearly identical to those of the
original haloes, but the core to virial radius ratio is reduced, with
the outer density profile of the merger remnant well approximately by
$\rho\propto r^{-3}$.  This result also holds for mergers with
parent halo mass ratios of 3:1, implying all major mergers of two
cored haloes result in cored haloes.

\item Mergers of a cuspy halo with either a cored halo or a second
cuspy halo result in a cuspy halo.  If the cuspy halo starts out with
an NFW form, the merged halo also has an NFW-like form, with a
slightly shallower inner cusp of $\rho\propto r^{-0.7}$ instead of
$r^{-1}$.

\item The particles from a given halo retain memory of their
progenitor's density profile.  In major mergers of a cored halo with a
cuspy halo, the particles originating from the cuspy halo end up with
a cuspy profile, and those from the cored halo end up with a cored profile.
(The cuspy particles dominate the total potential at the centre of the
merged halo, so the resulting profile is cuspy.)  The core
particles in the inner 10 $h^{-1}$ kpc of the remnant are
significantly heated.

\item The merger remnants relax from the inside outwards, attaining
Gaussian velocity distributions near their centres but not throughout
the halo.  Violent relaxation due to time-varying potential is
effective only during the initial phase of the mergers; phase mixing
is likely the dominant relaxation process at late times.

\item Including a moderate amount of orbital angular momentum in the
form of a non-zero impact parameter has little discernible effect on
our findings.
\end{enumerate}

The set of simulations presented in this paper has been performed for
mergers of non-spinning haloes in a set bound orbit.  This
clearly only probes a restricted range of parameter space.  A possible
area for future investigations is to include non-zero initial spins,
which can influence tidal interactions significantly but may only
affect the final phase space structure weakly \citep{white79}.  We
have also chosen not to convolute the issues in this paper by ignoring
minor mergers with satellite haloes and continuous accretion of diffuse
material.  The latter mechanisms may be responsible for generating the
first generation of cuspy haloes from initially cored haloes, a process
not achieved by major mergers of cored haloes according to our study.

Although the length and time scales quoted in this paper are for
Milky-Way sized haloes of $10^{12} M_{\odot}$, there is nothing in
the numerical code or in our analysis that is special to this
particular mass scale (except a mildly mass-dependent concentration
parameter for NFW haloes).  Our conclusions in this paper are thus
likely to be applicable to major mergers of comparable-mass haloes
over a wide mass range. Our finding that cuspy haloes are resilient to
major mergers hints that the existence of cuspy haloes is natural in
the hierarchical structure formation scenario: the build-up to larger
mass haloes via repeated merging may be sufficient to produce cuspy
haloes from cored haloes and preserve the general structure of cuspy
haloes. 

\section*{Acknowledgments} 

We thank Jon Arons, Andrew Benson, Andrey Kravtsov, and Volker
Springel for useful discussions and Jeff Filippini for help with the
energy analysis.  This research used resources of the National Energy
Research Scientific Computing Center, which is supported by the Office
of Science of the U.S. Department of Energy under Contract
No. DE-AC03-76SF00098.  C.-P. M is partially supported by an Alfred
P. Sloan Fellowship, a Cottrell Scholars Award from the Research
Corporation, and NASA grant NAG5-12173.

\label{lastpage}

\end{document}